# Extraction of the Anomalous Nernst Effect in the Electric Measurement of the Spin Orbit Torque

M. Kawaguchi[1], T. Moriyama[1], H. Mizuno[1], K. Yamada[1], H. Kakizakai[1], T. Koyama[2], D. Chiba[2], T. Ono [1]

1. ICR, Kyoto University, Kyoto 611-0011, Japan
2. Department of Applied Physics, The University of Tokyo, Tokyo 113-8656, Japan

Abstract

Spin orbit torque has been intensively investigated because of its high energy efficiency in manipulating a magnetization. Although various methods for measuring the spin orbit torque have been developed so far, the measurement results often show inconsistency among the methods, implying that an electromotive force, such as Nernst effect, irrelevant to the spin orbit torque may affect the measurement results as an artifact. In this letter, we developed a unique method to distinguish the spin orbit torque and the anomalous Nernst effect. The measurement results show that the spin orbit torque can be underestimated up to 50% under the influence of the anomalous Nernst effect.

Spin orbit torque caused by Rashba effect[1] and spin Hall effect[2, 3] has been attracting a great deal of attention in both physical and application viewpoints. It is generally observed in ferromagnetic multilayers having asymmetric layer structures composed of heavy metal layers and ferromagnetic layers. Since the spin orbit torque provides a higher conversion efficiency from the electric current to the spin torque, comparing to the classical Oersted field, it would be a key technology for future low energy consumption spintronic devices[4, 5, 6]. Although physical phenomenologies of both Rashba effect and the spin Hall effect are theoretically well understood[1, 2, 3], there have been a debate on how those effects arise in the actual heavy metal/ferromagnetic multilayer samples. The uncertainties of the experimental results as well as the inconsistent results between measurement techniques have been greatly interrupting the progress of the physical understanding of the spin orbit torque.

For experimentally determining the spin orbit torque, various techniques have been employed so far, including spin torque ferromagnetic resonance (ST-FMR)[7], magnetization switching[8], domain wall dynamics[9], second harmonic measurement[10], and dc Hall measurement[11]. Each technique has drawbacks and advantages as well as a limited capability. Which technique one should use is depending on the sample materials and sample structures. Besides, one of the biggest issues is that the final results are often varied with the measurement techniques[4, 12]. This, in other words, means that the physical assumptions and principles behind the measurement techniques are not somewhat sufficiently taken into account. The above all measurement techniques are based on the electrical measurement which relies on the magnetoresistance of the ferromagnetic materials to sense the magnetization response to the spin torque. A major issue on this sort of measurement is that it is prone to artifacts by various electromotive forces that are irrelevant to the magnetoresistance[13, 14]. In particular, the anomalous Nernst effect (ANE) due to the temperature gradient induced in the multilayer structure by the Joule heating has been known to contribute to the artifacts in some of the measurement techniques. The influences of the ANE are often mathematically characterized to be insignificant in the characterization of the spin orbit torque[15] as it is quite hard to experimentally characterize the temperature gradient.

In this paper, it will be shown that, within the framework of our dc Hall measurement technique, it is possible to experimentally distinguish the ANE contribution by a simple set of measurements. We show that the ANE can be quantitatively characterized by analyzing an external field strength dependence of Hall voltages. We found that the ANE can affect the accuracy of the spin orbit torque estimation by up to 50%.

Figure 1 shows the experimental setup for the dc Hall measurement. We used a multilayer of Ta(2.5 nm)/ Pt(1.7 nm)/ Fe(0.6 nm)/ MgO(2.2 nm) as a model system. The multilayer is deposited on a GaAs (001) substrate by r.f. magnetron sputtering. The Hall bar structure with a 30 μm wide channel is fabricated by photolithography and Ar ion milling. The Hall measurements with rotating external fields are curried out to determine the spin orbit torque and ANE voltages induced by the temperature gradient. The relative angle between $H_{ex}$ and x-axis is defined as $\theta$. The Hall voltages are measured at various $\theta$ by changing the polarity of the current. All measurements are carried out at 310 K.

In the static state, the spin orbit torque can be interpreted as a current induced effective field. There are two different effective fields corresponding to two different spin orbit torques, which are called damping like torque and field like torque. These two effective fields can be determined by the Hall measurements because the magnetization tilting due to the effective fields induces Hall resistance changes by the planar Hall effect and the anomalous Hall effect[11]. Therefore, the magnetization response to these effective fields is reflected in the Hall voltage.

The Hall resistance differences between two current polarities are carefully examined to exclude the resistance changes irrelevant to the effective fields. Figure 2 shows the experimental data of the Hall resistance difference, $R_{\text{diff}} \equiv R_{\text{Hall}}(+I) - R_{\text{Hall}}(-I)$, as a function of $\theta$. The resistance difference $R_{\text{diff}}$ is useful for analyzing effective fields induced by the current. The two effective fields are determined by $R_{\text{diff}}$. In a sample having an in-plane magnetization, the spin orbit torque induces effective fields in two different directions; the transverse field $H_T$ and the perpendicular field $H_P$ (as shown in Fig. 1). The counterparts of $H_T$ and $H_P$ are the field like torque and the damping like torque, respectively. $H_T$ and $H_P$ modify $R_{\text{diff}}$ differently as shown below.

$H_T$ and $H_P$ independently modify the Hall voltage by the planar Hall effect and the anomalous Hall effect[11]. When the ANE voltage is in account, three different sources of voltage must be considered. In the following paragraphs, how these voltages arise is discussed in details.

The ANE is a thermoelectric phenomenon that induces a voltage in the presence of a magnetization and temperature gradient.[13, 16] A temperature gradient $\nabla T$ drives electron flows and the magnetization $m$ bends the electron trajectories. As a result, the voltage in the direction of $\nabla T \times m$ is induced. Because a voltage projected on y-axis is detected as a Hall voltage in our setup, the voltage contribution of ANE, $V_{\text{ANE}}$, is proportional to $(\nabla T \times m)_y$. This relation indicates that, in an in-plane magnetization system, $V_{\text{ANE}}$ is proportional to $\cos\varphi$, where $\varphi$ is the relative angle between magnetization and current. The current dependence of $V_{\text{ANE}}$ is also important to distinguish $V_{\text{ANE}}$ from the voltage due to the conventional Hall effects. The magnitude of $V_{\text{ANE}}$ is proportional to $\nabla T$ which drives electrons. The temperature gradient $\nabla T$ is inevitably induced by a current flow in order to carry out the Hall measurements. The current induced $\nabla T$ should have a linear relationship with the Joule heating power in a steady state because of the Fourier's law indicating that the heat flow that compensate heating power is proportional to $\nabla T$. Therefore, the magnitude of $V_{ANE}$ is proportional to $I^2$. After all, the ANE voltage $V_{ANE}$ is proportional to both $\cos\varphi$ and $I^2$ and the contribution of the ANE to the $R_{diff}$ is represented as,

$$R_{ANE}(\cos(\varphi(+I)) - \cos(\varphi(-I))) \sim 2R_{ANE}\cos\theta. \ (\varphi \sim \theta) , \quad (1)$$

where $R_{ANE}$ is the $V_{ANE}/I$ at $\varphi = 0°$.

The effective field $H_P$ is caused by an injected spin from Pt to Fe layer due to the spin Hall effect. A spin angular momentum transfer from injected spin exerts a torque on the magnetization. This torque is considered as an effective field in a static state. The torque exerting on a magnetization $m$ is written as $\frac{\hbar}{2eM_st}J_s(m \times (\sigma \times m))$,[17] $J_s$ is the spin current density, $\sigma$ is an electrons spin vector, $\hbar$ is the Planck constant, $e$ is the elementary charge, $M_S$ is the saturation magnetization, and $t$ is the ferromagnetic layer thickness. $H_P$ is equal to $-\frac{\hbar}{2eM_st}J_s(\sigma \times m)$. The injected spin vector $\sigma$ is along $y$-axis in our experimental setup because the injected spin is aligned with the direction of $J \times J_s$, where $J$ is the current density vector. The magnitude of $\sigma \times m$ is proportional to $\cos\varphi$ in this situation. As a result, the magnitude of $H_P$ is proportional to $\cos\varphi$. The $H_P$ makes the magnetization rotation out of the sample plane and $m_z$, the $z$ component of $m$, induces a Hall voltage due to the anomalous Hall effect. The magnitude of $m_z$, or the Hall voltage, is simply proportional to $\cos\varphi$ when the $H_P$ is sufficiently small comparing to the external in-plane field $H_{ex}$. The $H_P$ contribution to $R_{diff}$ is written as

$$R_{HP}(\cos(\varphi(+I)) - \cos(\varphi(-I))) \approx 2R_{HP}\cos\theta, \quad (2)$$

where $R_{HP}$ is the Hall resistance induced by $H_P$ at $\theta = 0°$.

The transverse effective field $H_T$ is the field induced orthogonal to the current. $H_T$ can be treated as an additional in-plane field. $H_T$ rotates the magnetization in the plane. This rotation can be detected by the planar Hall effect. Therefore, $H_T$ can be determined by the planar Hall effect. The magnetization is aligned with the total field $H_{total}$ that is sum of $H_T$ and the external field $H_{ex}$. The $x$ and $y$ components of $H_{total}$ are $H_{ex}\cos\theta$ and $H_{ex}\sin\theta + H_T$, respectively. The magnetization angle $\varphi$ is calculated from these components as $\varphi = \text{Arctan}(\frac{\sin\theta+\delta}{\cos\theta})$, where $\delta = H_T/H_{ex}$. $H_T$ contribution to $R_{diff}$ is written as,

$$R_{PH}(\varphi(+I)) - R_{PH}(\varphi(-I)) = R_{PH}(\text{Arctan}(\frac{\sin\theta+\delta}{\cos\theta})) - R_{PH}(\text{Arctan}(\frac{\sin\theta-\delta}{\cos\theta})) \quad (3)$$

where $R_{PH}(\varphi)$ is a function of the planar Hall resistance that is proportional to $\sin 2\varphi$.

Finally, the Hall resistance difference $R_{diff}$ is written as,

$$R_{diff} = 2R_{cos}\cos\theta + R_{PH}(\text{Arctan}(\frac{\sin\theta+\delta}{\cos\theta})) - R_{PH}(\text{Arctan}(\frac{\sin\theta-\delta}{\cos\theta})) \quad (4)$$

where $R_{cos} \equiv R_{HP} + R_{ANE}$.

The data shown in Fig. 2 are well fitted by Eq. 4 with a resistance offset and angle offset. The values of $H_T$ and $R_{cos}$ are determined for various current by the fitting. The results are shown in Fig. 3 as a function of the current $I$. $H_T$ show the linear relation with current $I$, which is expected in the framework of the spin orbit torque[18]. As $R_{cos}$ contains the information of the ANE, we will focus on $R_{cos}$ vs. $I$ below in more detail.

$R_{ANE}$ is defined as $V_{ANE}/I$ and the $V_{ANE}$ is proportional to $I^2$ because the current heating power is proportional to $I^2$. As a result, the $R_{ANE}$ is proportional to $I$. The Hall resistance induced by perpendicular *effective* field $H_P$, $R_{HP}$, is represented as $\frac{dR_{AH}}{dH_{perp}}H_P$ where $\frac{dR_{AH}}{dH_{perp}}$ is the Hall resistance change induced by the anomalous Hall effect with respect to the externally applied perpendicular field. $\frac{dR_{AH}}{dH_{perp}}$ is independent of $H_P$ when $H_P$ is sufficiently small comparing to the demagnetizing field. $H_P$ is indeed small enough in our experimental condition. As $H_P$ is proportional to the current, the resistance $R_{HP}$ is linearly proportional to the current.

The value of $\frac{R_{cos}}{I}$ is obtained from the linear fitting of $R_{cos}$ against the current as shown in Fig. 3. $R_{\text{ANE}}$ and $R_{\text{HP}}$ must be distinguished in order to determine $H_\text{P}$. $R_{\text{ANE}}$ and $R_{\text{HP}}$ have different external field dependence. While $R_{\text{ANE}}$ is insensitive to the in-plane magnetic field, $R_{\text{HP}}$ is sensitive to the in-plane field. $R_{\text{HP}}$ is due to the anomalous Hall effect and, therefore, the perpendicular component of magnetization is important. On the other hand, $R_{\text{ANE}}$ is induced by the anomalous Nernst effect and in-plane component of the magnetization is important. The magnetization should point to the direction of $H_{\text{ex}} + H_\text{P}$ in the simple model if the transverse field is insignificant. The perpendicular and in-plane components of the magnetization are represented as $H_\text{P} M_S/\sqrt{H_{\text{ex}}^2 + H_P^2}$ and $H_{\text{ex}} M_S/\sqrt{H_{\text{ex}}^2 + H_P^2}$, respectively. The external field dependence of the magnetization components are $\propto 1/H_{\text{ex}}$ (perpendicular) and $\propto 1 - H_P^2/2H_{\text{ex}}^2$ (in-plane), where $H_{\text{ex}} \gg H_\text{P}$. These relations indicate that $R_{\text{ANE}}$ does not depend on the external field strength while $R_{\text{HP}}$ does (when the external field is strong enough).

The value $\frac{R_{\cos}}{I}$ is written as

$$\frac{R_{\cos}}{I} = \frac{R_{\text{HP}} + R_{\text{ANE}}}{I} = \frac{H_\text{P}}{I} \cdot \frac{dR_{\text{AH}}}{dH_{\text{perp}}} + \frac{V_{\text{ANE}}}{I^2}. \quad (5)$$

This equation shows that $\frac{R_{\cos}}{I}$ is represented as a linear function of $\frac{dR_{\text{AH}}}{dH_{\text{perp}}}$ because $\frac{V_{\text{ANE}}}{I^2}$ and $\frac{H_\text{P}}{I}$ are independent of the current and external field strength. Therefore, $\frac{R_{\cos}}{I}$ and $\frac{dR_{\text{AH}}}{dH_{\text{perp}}}$ depends on the external field strength. Values for $\frac{R_{\cos}}{I}$ and $\frac{dR_{\text{AH}}}{dH_{\text{perp}}}$ are obtained from the Hall measurements with various external field strengths. Figure 4 shows the experimental results of $\frac{R_{\cos}}{I}$ and $\frac{dR_{\text{AH}}}{dH_{\text{perp}}}$ at $\mu_0 H_{\text{ex}} = 300$ mT~475 mT (25 mT steps). The experimental data are well fitted by Eq. 5 as shown in Fig. 5. Non-zero intercept of the fitting line at y-axis indicates that anomalous Nernst voltage is induced in the sample. The value of $\frac{V_{\text{ANE}}}{I^2}$ and $\frac{H_\text{P}}{I}$ are obtained from the linear fitting. The ratio of the Nernst voltage against the voltage induced by the perpendicular effective field is 9 % at $\mu_0 H_{\text{ex}} = 400$ mT. The spin Hall angle, $J_\text{s}/J_{\text{Pt}}$, is calculated from the equation $H_P = \hbar J_S / 2eM_S t$, now $J_{Pt}$ is the current density in the Pt layer. The calculated value of $J_s/J_{Pt}$ is 0.12, which is in good agreement with previous reports.[7, 19, 20] if $V_{ANE}$ were not to be taken into account, $J_s/J_{Pt}$ would be estimated to be 0.11. The result clearly shows that the ANE indeed gives a significant correction in the spin Hall angle estimation.

Finally, we applied the same method to the Fe/Pt multilayers having various Fe layer thicknesses. Figure 5 shows the experimental results of $J_s/J_{Pt}$ as a function of Fe layer thickness. All samples have structures of Ta(2.5 nm)/ Pt(1.7 nm)/ Fe($X$ nm)/ MgO(2.2 nm) . We found that $J_s/J_{Pt}$ with ANE correction is always greater than the one without the ANE correction. The ANE correction influences the accuracy of the $J_s/J_{Pt}$ up to 50 %. These results indicate that the ANE is not trivial in our dc Hall measurement and should be treated with care to obtain the accurate $J_s/J_{Pt}$.

In summary, we investigate how the ANE influences on our dc Hall measurement technique for measuring the effective fields. We show that the ANE can be distinguished from the Hall voltages induced by the effective fields by taking advantage of the difference in the external field and current dependence of these voltages. It is found that the ANE correction can influence the accuracy of the effective field estimation up to 50%. The results indicate that the ANE is non-negligible and must be taken into account to obtain correct results of the current induced effective field.

This work was supported by JSPS KAKENHI Grant Numbers 25・4681, 26870300, 15H05702, and 25220604.

Figure captions

Figure 1 Schematic illustration of the experimental setup. A current is applied along *x* axis. (a) The current induces Joule heating in the Pt layer (thicker and higher conductive than Fe layer) and the heat diffuses toward the Fe layer. (b) The current also gives rise to the spin orbit torque that induces the effective fields at the interface of Pt/Fe. (c) The Hall measurements are carried out with a rotating external magnetic field. The angle between the current and external field is $\theta$.

Figure 2 $R_{\text{diff}}$ as a function of the field angle $\theta$. The experimental data are shown in dots and the fitting curves by Eq. 4 are in lines.

Figure 3 (a) $R_{\cos}$ as a function of the current. (b) $\mu_0 H_T$ as a function of the current. These values are obtained from fitting results in Fig. 2. The values are linearly proportional to the current.

Figure 4 $R_{\cos}$ as a function of $dR_{\text{AH}}/dH_{\text{perp}}$. The data points are well fitted by linear function shown in Eq. 5. The slope indicates the perpendicular effective field. The non-zero interception indicates that there is a significant ANE in the Hall measurements.

Figure 5 Comparison between the spin Hall angle estimation with and without ANE correction. The red squares represent $J_S/J_{\text{Pt}}$ with the ANE correction and the blue squares without the correction. The values after the correction are always greater than the one without. The difference between the values before and after the correction varies from 9% to 50%.

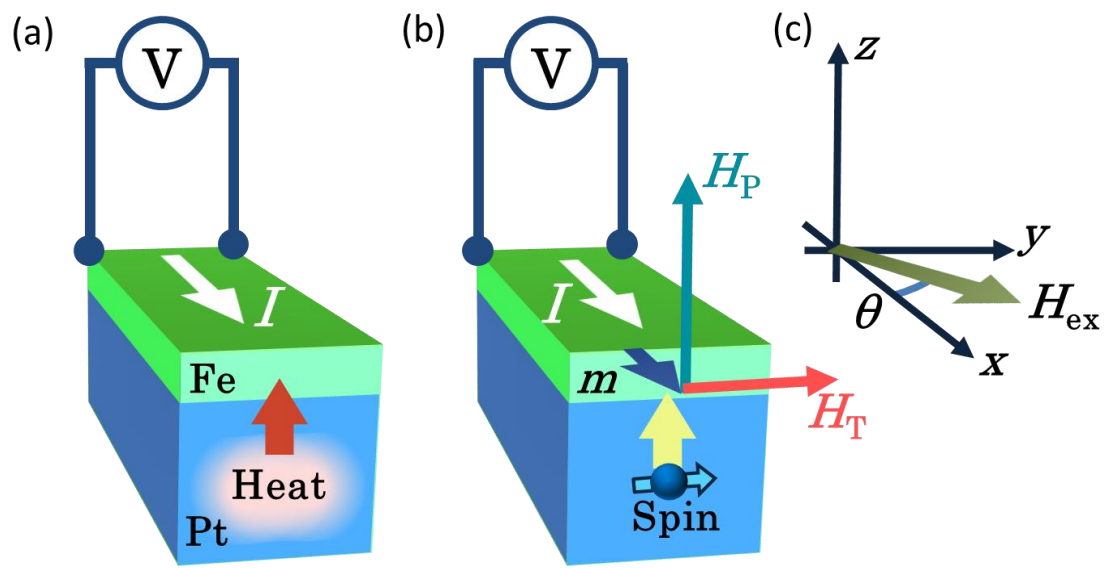

Figure 1

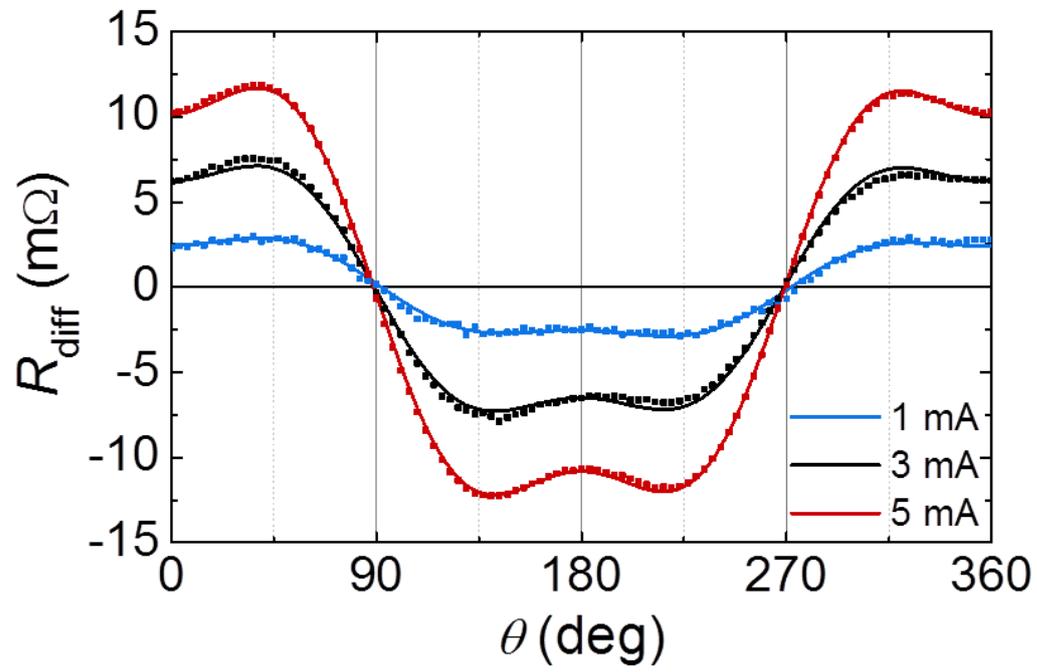

Figure 2

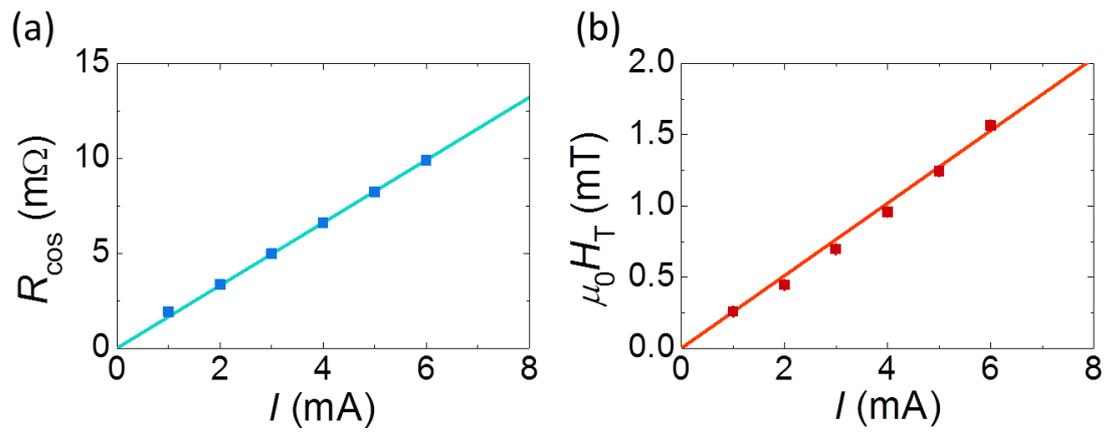

Figure 3

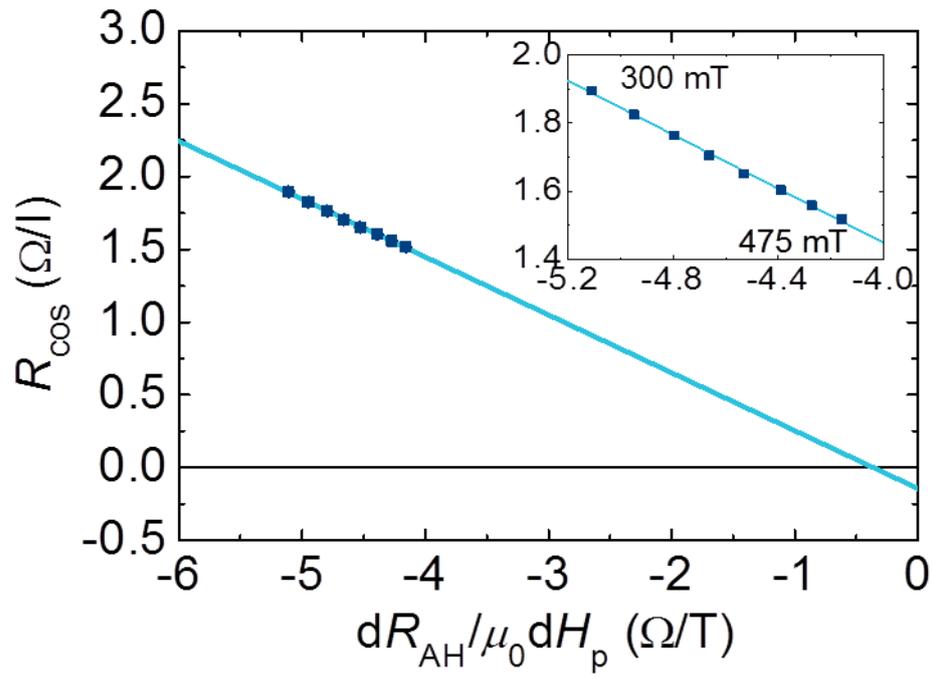

Figure 4

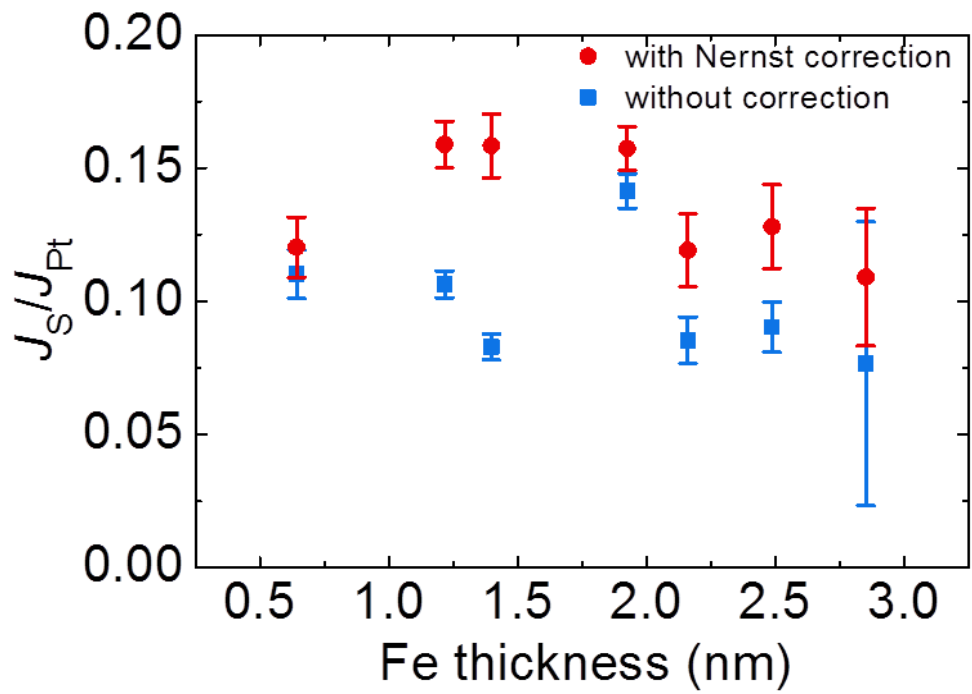

Figure 5